# Flexible and Efficient Semi-Empirical DFTB Parameters for Electronic Structure Prediction of 3D, 2D Iodide Perovskites and Heterostructures

*Junke Jiang[1], Tammo van der Heide[2], Simon Thébaud[1,*], Carlos Raúl Lien-Medrano[2], Arnaud Fihey[3], Laurent Pedesseau[1], Claudio Quarti[4], Marios Zacharias[1], George Volonakis[3], Mikael Kepenekian[3], Bálint Aradi[2], Michael A. Sentef[2], Jacky Even[1,*], Claudine Katan[3,*]*

[1]Univ Rennes, INSA Rennes, CNRS, Institut FOTON - UMR 6082, F-35000 Rennes, France

[2]Institute for Theoretical Physics and Bremen Center for Computational Materials Science, University of Bremen, 28359 Bremen, Germany

[3]Univ Rennes, ENSCR, CNRS, ISCR - UMR 6226, F-35000 Rennes, France

[4]Univ Laboratory for Chemistry of Novel Materials, Materials Research Institute, University of Mons, Place du Parc 20, Mons 7000, Belgium

KEYWORDS: Iodide perovskites, 2D perovskites, DFTB, electronic structure, band gap, effective masses

## ABSTRACT

Density Functional Tight-Binding (DFTB), an approximative approach derived from Density Functional Theory (DFT), has the potential to pave the way for simulations of large periodic or non-periodic systems. We have specifically tailored DFTB parameters to enhance the accuracy of electronic band gap calculations in both 3D and 2D lead-iodide perovskites, at a significantly reduced computational cost relative to state-of-the-art *ab initio* calculations. Our electronic DFTB parameters allow computing not only the band gap but also effective masses of perovskite materials with reasonable accuracy compared to existing experimental data and state-of-the-art DFT calculations. The electronic band structures of vacancy-ordered and, lead- and iodide-deficient perovskites are also explored. Additionally, we demonstrate the efficiency of DFTB in computing electronic band alignments in perovskite heterostructures. The DFTB-based approach is anticipated to be beneficial for studying large-scale systems such as heterostructures and nanocrystals.

## INTRODUCTION

Metal halide perovskites (MHPs) have emerged as promising materials for various optoelectronic applications, including solar cells [1-3], light-emitting diodes [4-6], and photodetectors [7,8]. Understanding the intrinsic electronic and optical properties is crucial for further advancing their performance and exploring new device designs. The lead iodide perovskites (LIPs) are extensively studied because their band gaps lie in the optimal range for photovoltaic applications [9]. While the highest solar to electricity conversion efficiencies have been achieved with 3D MHPs, their 2D counterparts offer various benefits and add features such as stability under operation, structural and chemical diversities, strongly bound excitons, anisotropy of the physical properties and various features related to quantum and dielectric confinement [10,11]. Additionally, the engineering of complex heterostructures combining 3D and 2D MHPs is currently a prominent technological trend [12,13]. The electronic band gap values and effective masses of both 3D and 2D compounds as well as the alignments of their band edges play a crucial role in their ability to be integrated effectively in the active zones of solar cells [13] or light-emitting diodes [14].

The electronic band gap is fundamentally defined as the difference between the ionization energy and the electron affinity, and is commonly measured using spectroscopic techniques like ultraviolet photoemission spectroscopy (UPS) for occupied states and inverse photoemission spectroscopy (IPES) for unoccupied states [15,16]. From the perspective of semiconductor physics, the band gap is expected to be related to the computed energy difference between the valence band maximum (VBM) and the conduction band minimum (CBM). But practically predicting band gaps can be challenging, and various methodologies have been developed for that purpose. The Kohn-Sham density functional theory (DFT) is a popular computational method used in the condensed matter community [17,18]. However, DFT tends to underestimate the electronic band gap due to self-interaction errors in the occupied states [19] and the absence of the derivative discontinuity [20,21]. To improve the prediction of electronic band gaps in solids, various methods have been developed, such as the GW approximation in many-body perturbation theory [22,23], hybrid functionals [24-27], the meta-GGA SCAN functional [28], DFT+U [29], and DFT-1/2 [30]. Many-body perturbation theory in the GW approximation is in principle the most accurate way to determine the fundamental band gap, however, it is challenging to apply to large systems (over hundreds of electrons) due to computational limitations. Even though the hybrid functionals are less demanding for large systems than GW, they still require significant computational resources and improved description of the dielectric screening function. In a previous work, some of us have demonstrated the accuracy of the DFT-1/2 method (Slater half occupation scheme), which comes at the cost of a DFT calculation for predicting the band gap, effective masses, and energy level alignments of 3D and 2D MHPs and its heterostructures [31]. However, the computational cost of standard DFT still restricts the system size, and additionally, DFT-1/2 requires a careful choice of predominant atomic orbitals in band-edge electronic states, i.e. those affected by the electronic population corrections [30]. Therefore, an efficient and computationally cost-effective band gap correction scheme is essential to handle systems approaching more realistic sizes, relevant for multilayers or nanostructures in real devices (typically tens of nanometers and more than one thousand valence electrons). Among the various computationally cost-effective quantum frameworks able to achieve this goal, Density Functional Tight-Binding (DFTB) emerges as a particularly relevant approach.

DFTB is an approximation to DFT that aims at combining the efficiency of Tight-Binding (TB) models with the accuracy of reference methods [32]. It offers a computationally efficient approach to studying large systems, striking a reasonable balance between accuracy and computational cost. In the context of conventional semiconductors, DFTB has been successfully employed to investigate the electronic properties of heterostructures, surfaces, atomic clusters and point defects [33-37]. It should be noted that the DFTB method requires an appropriate parametrization that accounts for the specific characteristics of the material under investigation, in order to achieve reliable and accurate results [38]. Parametrizations for DFTB usually aim at replicating DFT ground state properties of materials. However, the presence of a few adjustable parameters in the DFTB framework provides further flexibility in tackling excited state problems, including the calculation of electronic band gaps. This opens the possibility for realistic electronic structure predictions of MHPs [39].

One major challenge in applying DFTB to MHPs stems from the absence of standard parametrizations for lead (Pb), a crucial element in these materials. In MHPs, Pb atoms form bonds with halide atoms and interact with organic cations within the perovskite structure. These interactions are complex and require meticulous parametrization. Addressing this problem properly is essential for enhancing the predictive power and applicability of DFTB in the study of MHPs. Recent efforts have been made to develop suitable Pb parameters for DFTB calculations. Yam *et al.* proposed a parametrization of Pb for the simulation of $MAPbI_3$ (where MA stands for methylammonium, with general formula $CH_3NH_3$) that yielded correct electronic band gap and effective mass predictions [39]. However, this work did not include any Cs parameters, an element widely used nowadays in 3D MHPs and alloys. The parameter transferability to other perovskites, lower dimensional materials, or heterostructures was not tested either. Nakai *et al.* introduced new parameters for Pb, I, and Cs and studied the photoexcited-state dynamics of $CsPbI_3$ and $MAPbI_3$ [40]. However, spin-orbit coupling (SOC), a crucial factor in accurately describing the electronic structure of MHPs [41], was not considered, resulting in a biased description such as incorrect degeneracy of the band edges. Therefore, a complete set of DFTB parameters including SOC that is transferable from hybrid to inorganic perovskites, as well as lower dimensional compounds, is still missing.

In this study, we present DFTB parametrizations tailored for LIPs to improve the predictions of the electronic structures of 3D compositions ($CsPbI_3$, $MAPbI_3$, and $FAPbI_3$ [FA = $CH(NH_2)_2$]), leading to a fair agreement with available experimental data. Moreover, we also present, to the best of our knowledge, the first application of DFTB to 2D MHPs. The DFTB method captures specific features of the electronic band structures of model 2D LIPs (e.g., the fictitious $Cs_2PbI_4$ compound, and the well-known $BA_2PbI_4$, and $PEA_2PbI_4$ layered perovskites), including quantum confinement effects and the anisotropy of various properties. We demonstrate good agreement between DFTB results and experimental electronic band gaps and reduced effective masses. To quantify this assessment, we compared our DFTB results to available experimental data. For the examined set of 3D and 2D perovskites, the mean absolute error (MAE) on computed band gaps is 0.09 eV for 3D and 0.24 eV for 2D. For reduced effective masses MAE amounts to 0.012 $m_0$ and 0.056 $m_0$, respectively. Furthermore, this first attempt beyond 3D MHPs holds promise for expanded applications to other low-dimensional perovskite nanostructures, such as vacancy-ordered perovskites and lead- and iodide-deficient perovskites. Additionally, the DFTB band alignments of 3D/2D perovskite heterostructures show reasonable accuracy, opening up the possibility for scaling up to device-size model tests in the future.

# METHODS

## I. Density Functional Tight-Binding (DFTB)

We present an overview of the DFTB theory to illustrate its capacity to meet the demands for efficient and transferable calculations in large and complex systems such as MHPs. We also discuss the parameters available within the theory. A more detailed exploration of the DFTB theory and parametrization procedure is available in the reference papers [38,42-44]. For the sake of clarity, this section provides equations for non-periodic systems. For detailed equations applicable to periodic systems, the reader may refer to the periodic DFTB theory paper [45].

The DFTB models are derived from Kohn-Sham (KS) DFT by expanding the total energy functional. Starting from a properly chosen reference density $\rho_0$, where $\rho_0$ is usually chosen as a sum of neutral or confined atomic densities, the ground state density $\rho(r)$ is then represented as:

$$\rho(r) = \rho_0(r) + \delta\rho(r) \tag{1}$$

The total energy expression then expands the KS energy functional in a Taylor series. In our study we use the Self-Consistent-Charge Density Functional Tight-Binding (SCC-DFTB) up to second order, which is also called DFTB2:

$$E[\rho_0 + \delta\rho] = E^0[\rho_0] + E^1[\rho_0, \delta\rho] + E^2[\rho_0, (\delta\rho)^2] \tag{2}$$

where $E^0[\rho_0]$ represents the repulsive term essential for determining atomic forces within a system, dependent solely on the reference density $\rho_0$. It can usually be approximated in practice as a sum of pair potentials:

$$E^0[\rho_0] \approx E_{\text{rep}} = \frac{1}{2}\sum_{ab} V_{ab}^{\text{rep}} \tag{3}$$

These potentials are typically determined through comparison with DFT calculations or are fitted to empirical data. In scenarios where the electronic structure calculations are conducted with a fixed atomic structure that is derived from experimental data, the $E^0[\rho_0]$ term becomes irrelevant.

$E^2[\rho_0, (\delta\rho)^2]$ is the second order charge fluctuations arising from Coulomb and exchange-correlation (XC) interactions. It is approximated by a Coulomb-like interaction in DFTB, consisting of a long range $1/r$ interaction and a short-range part parameterized by ab initio atomic parameters, like the atomic Hubbard U value.

$E^1[\rho_0, \delta\rho]$ is the electronic structure term, which dominates the electronic structure and arises from a Hamiltonian built on the reference density $\rho_0$, usually written as:

$$E^1[\rho_0, \delta\rho] = \sum_i n_i \langle \psi_i | \hat{H}[\rho_0] | \psi_i \rangle = \sum_i \sum_\mu \sum_\nu n_i c_\mu^i c_\nu^i H_{\mu\nu}^0 \tag{4}$$

where the $|\psi_i\rangle$ are electronic eigenstates, $n_i$ their population and $c_\mu^i$ their coefficients on a localized orbital basis (see below). The so-called non-SCC Hamiltonian $H_{\mu\nu}^0$ is determined by (precalculated) two-center integrals and atomic orbital energies, obtained from *ab initio* calculations.

First, atomic parameters such as the energies of s and p orbitals, the Hubbard parameter U, and optionally, the spin-coupling parameters W of isolated atoms are calculated from self-consistent all-electron DFT calculations.

Next, the electron density $\rho_0$ of isolated atoms is evaluated by means of the equation below:

$$\left(T + V_{\mathrm{nuc}} + V_{\mathrm{Har}} + V_{\mathrm{XC}} + \left(\frac{r}{r_{\mathrm{dens}}}\right)^n\right) \phi_i = \varepsilon_i \phi_i \tag{5}$$

The Hamiltonian contains the kinetic energy $T$, the electron-nucleus attraction $V_{nuc}$, the Hartree energy $V_{Har}$, the XC contribution $V_{XC}$, and an additional confining potential of the form $(\frac{r}{r_{dens}})^n$ that significantly enhances the accuracy of the DFTB method, where $r_{dens}$ represents the compression radius for the atomic density.

The starting wave-function and Hamiltonian of the system are determined following two major approximations:

i) a valence-only minimal basis set ($\eta_i$) yielding linear combinations of atomic orbitals (LCAO) for the eigenstates $|\psi_i\rangle$,

$$|\psi_i\rangle = \sum_{\mu} c_{i\mu} |\eta_\mu\rangle \tag{6}$$

These basis functions $\eta_\mu$ are calculated for isolated atoms, solving a Schrödinger equation with a different (shell-resolved) compression radii $r_{\mathrm{wf}}$:

$$\left(T + V_{\mathrm{nuc}} + V_{\mathrm{Har}} + V_{\mathrm{XC}} + \left(\frac{r}{r_{\mathrm{wf}}}\right)^n\right) \eta_\mu = \varepsilon_\mu \eta_\mu \tag{7}$$

ii) a two-center approximation to the Hamiltonian operator $\hat{H}[\rho_0]$. The Hamiltonian matrix elements $H^0_{\mu\nu}$ are precalculated with a DFT program and are tabulated.

$$H^0_{\mu\nu} = \langle \eta_\mu | \hat{H}[\rho_0] | \eta_\nu \rangle \tag{8}$$

These two-center Slater-Koster parameters for functions $\mu$ and $v$ located on different atoms $a$ and $b$, respectively, are computed numerically as:

$$H_{\mu\nu}^{0} = \int \eta_{\mu}(\widehat{H}[\rho_a + \rho_b])\eta_{\nu} dV \tag{9}$$

where the Hamiltonian is constructed using the superposition of atomic densities as the starting point for the SCC interactions, which leads to converged densities and occupations in $H_{\mu\nu}$. In fact, there are numerous self-consistent calculations on molecules and solids showing that the electron densities can be approximated as a superposition of compressed atomic densities [42].

## II. Parametrization of DFTB

In this work we focus on the electronic structure and parametrize only the electronic part of the DFTB energy (we use fixed atomic positions, thus ignoring the repulsive potentials in $E^0$). In Equations (5, 7), the confining potentials are related to the DFTB parameters $r_{\text{dens}}$, $r_{\text{wf}}$ and $n$. These parameters must be determined per chemical element. For the problem at hand, we need parameters for Pb, I, Cs, C, N, H. We adopt the parameters of H, C, and N from the set "*mio-1-1*" [43,46] and the parameter of Cs are taken from Ref. [47]. For the Pb and I elements, which dominate the electronic structure close to the band edges [41,48], new parameters are proposed. The atomic calculations for parametrization are done at the all-electron DFT level, with PBE functional and scalar relativistic effects of the zeroth-order regular approximation (ZORA) Hamiltonian [49,50], using the parametrization tool skprogs [51]. We use the density superposition scheme as described above and in Ref. [43], which allows distinct compression radii per element. For optimizing $r_{\text{dens}}$, $r_{\text{wf}}$, and $n$, we aimed at obtaining accurate band gap and effective masses at the VBM and CBM, which is essential for the performance of perovskites photovoltaic and electronic devices. We used a minimal valence-only basis $5s^2 5p^5$ for I and $6s^2 6p^2$ for Pb in DFTB. Further addition of 4d and 5d orbitals for I and Pb, respectively, yields essentially non-interacting states far from the VBM and CBM, and increases considerably the DFTB computational cost due to more than doubling the number of orbitals [39]. We also consider two other factors: i) SOC splitting, which is the energy difference between the CBM and higher energy CB levels that is an essential factor to properly evaluate photoabsorption spectra [41] and ii) the energy difference between the VBM and other VB levels at lower energies.

To accurately describe the band edge states of LIPs formed mostly from Pb and I atomic states, we employed a semi-empirical parametrization scheme. A DFTB parametrization procedure indeed involves multiple steps, typically including a fitting process to match reference DFT calculations. However, this standard approach usually provides electronic gap values at the DFT-level, which are underestimated by comparison to experiment. To find a DFTB parametrization more suitable for the description of MHP optoelectronic properties, we instead primarily targeted experimental quantities such as electronic gap, effective masses, and SOC splitting. According to the National Institute of Standards and Technology (NIST) Atomic Spectra Database, considering the +2 oxidation state typically expected of metals or metalloids, the SOC splitting for p-like orbitals of total angular momentum J = 3/2 and J = 1/2 for lead are set as 1.65 eV. Regarding halides with an $ns^2 np^5$ electronic configuration, the SOC splitting is also notable, amounting to 0.94 eV for iodine [52,53]. Additionally, for carbon, a published value of 0.009 eV is set [52-54], to keep consistency with earlier reports. In the context of the present work, spin-orbit effects is neglected for cesium, as it does essentially counterbalance the charge neutrality and only s-orbitals are included in the parametrization. We further neglected SOC for N, and H. The above mentioned fixed tabulated values for SOC splitting are used in all our DFTB calculations.

The chosen crystallographic structure to determine the DFTB electronic parameters for LIPs is the orthorhombic photoactive perovskite phase (space group Pnma) of the fully inorganic compound $CsPbI_3$ [55]. This black γ-phase can be undercooled below its transition temperature and stabilized as a metastable polytype at room temperature [56]. γ-$CsPbI_3$ is selected because it features a clear and pronounced octahedral tilting, ubiquitous in the room-temperature disordered polymorphous phases of hybrid organic-inorganic perovskites but not easily described at the standard DFT level [57-59]. Such features are critical for accurately modeling the electronic properties of MHPs under operational conditions, thereby providing a realistic basis for the simulations. It is important to reiterate that in our parametrization procedure, the reference is first and foremost the experiment, which gives us well-defined values for the gap and reduced effective mass of γ-$CsPbI_3$. To further describe additional features of the electronic band structure, such as the effective mass of electrons and holes in different directions, we have to rely on comparisons with high-level theoretical calculations, such as GW calculations [55]. This choice is justified in many cases by its reasonable agreement with experimental results and its ability to provide a more detailed representation of the electronic band structure. Upon completing the parametrization procedure with all constraints listed above, the optimized compression radii for Pb, I, and other elements from the database are displayed in **Table S1**.

## III. Computational accuracy and efficiency

To assess the accuracy of our as-established DFTB parameters with respect to the targeted $CsPbI_3$ electronic structure, **Figure 1** compares the band structure obtained from our DFTB parametrization with that from a self-consistent scissor GW (ss-GW) calculation as outlined in Ref. [55]. The ss-GW method, which iteratively applies a rigid scissor shift to the energies in both the Green's function and the screened Coulomb interaction, has proven effective in predicting the quasi-particle band gap and the reduced effective mass for γ-$CsPbI_3$ [55] and γ-$MAPbI_3$ [60]. The comparison reveals good agreement between DFTB and ss-GW results. For the energy difference between the VBM and the lower lying bands, we note a discrepancy of 0.17 eV, and for the spin-orbit splitting in the CB, perfect agreement is observed. Further analysis is provided in **Table 1**, where we compare the effective masses derived from both band structures (additional details on the effective masses along each direction can be found in **Figure S1-S2** and **Table S2-S3**). Importantly, the band gap of the room temperature γ-$CsPbI_3$ is found to be 1.70 eV in good agreement with experimental values [61,62]. The average effective masses over multiple paths in the first Brillouin zone for the orthorhombic γ-$CsPbI_3$ are 0.22 $m_0$ for electrons and 0.27 $m_0$ for holes. The reduced effective mass $\mu$ is 0.12 $m_0$, which agrees well compared to the experimental value of 0.114 $m_0$ [61]. Meanwhile, the effective masses calculated via DFTB align nicely with the values obtained from ss-GW. Therefore, the electronic structure of γ-$CsPbI_3$ calculated by our DFTB parameters closely matches the experiments and high-level ss-GW calculations.

We also tested the ability of DFTB to handle larger systems for electronic structure calculations, including models with over thousands of atoms including SOC. The cubic phase α-$CsPbI_3$, with an experimental lattice constant of 6.30 Å [56], serves as a building block for such large systems to compare the scaling-up behavior of DFTB in comparison to DFT (GGA-PBE) using the SIESTA package; details can be found in Calculation details section. As the structures reach several hundreds to thousands of atoms, the computational time for DFTB is 3 orders of magnitude less than that for DFT, as depicted in **Figure S5**. Importantly, our DFTB parameters are empirically fitted to experimental values for properties such as the band gap and effective masses in bulk materials. By accurately capturing these fundamental properties, the parameters are also expected to provide reliable predictions of quantum confinement effects in nanostructures.

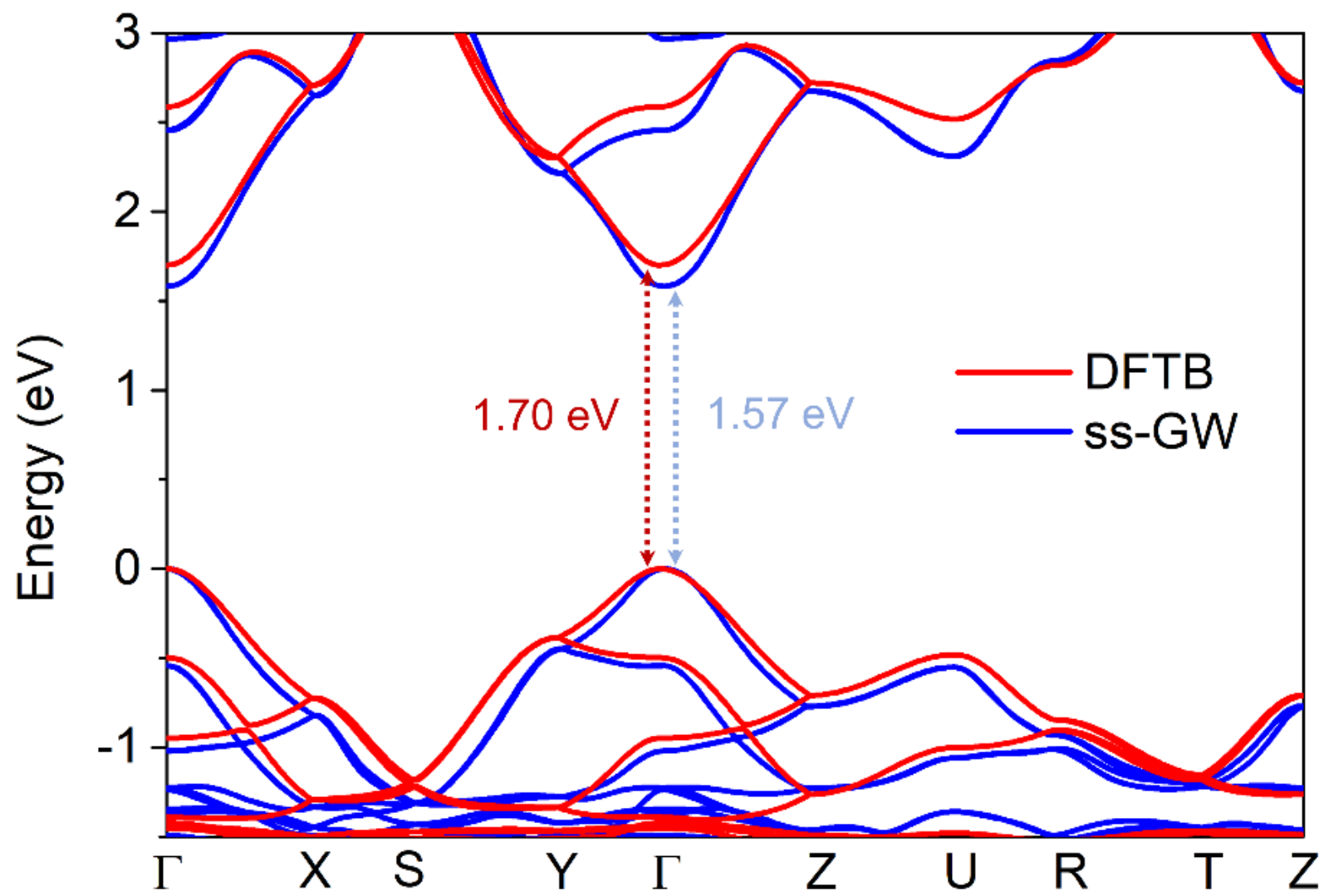

Figure 1. Band structure of γ-$CsPbI_3$ calculated using DFTB, compared with that obtained using self-consistent scissor GW (ss-GW) in Ref. [55]. SOC is accounted for both calculations.

Table 1. Band gap and effective masses of γ-$CsPbI_3$ calculated using DFTB compared to ss-GW values (Ref. [55]) and experiment (from Ref. [61]). $E_g$, $m_e$, $m_h$, and $\mu$ are the band gap, electron, hole, and reduced effective masses, respectively, where $\mu = m_e m_h/(m_e + m_h)$, given in units of the electron rest mass.

| | DFTB | ss-GW | Exp. |
|---|---|---|---|
| $E_g$ (eV) | 1.70 | 1.57 | 1.72 |
| $m_e$ ($m_0$) | 0.22 | 0.22 | - |
| $m_h$ ($m_0$) | 0.27 | 0.23 | - |
| $\mu$ ($m_0$) | 0.12 | 0.11 | 0.114 |

## IV. Calculation details

The electronic structure calculations at the DFT level are performed by using the SIESTA package with a basis set of finite-range numerical atomic orbitals [63]. Norm-conserving Troullier-Martins pseudopotentials are used for each atomic species to account for the core electrons [64]. $1s^1$, $2s^2 2p^2$, $2s^2 2p^3$, $6s^1$, $5s^2 5p^5$, and $6s^2 6p^2$ were used as valence electrons for H, C, N, Cs, I, and Pb, respectively. A double zeta polarization (DZP) basis set with an energy shift of 200 meV and a real space mesh grid energy cutoff of 300 Ry are used for all calculations. The GGA-PBE functional, combined with SOC in the on-site approximation, is used for the calculations [65]. DFTB electronic structure calculations are conducted with the DFTB+ software package [38], including SOC [66]. The GGA-PBE level of electronic structure calculations is used as a reference to directly compare the band structure shape with DFTB. All the effective masses are calculated using parabolic fitting [67] of the CBM and VBM curvatures from the Γ point along the x, y, and z directions. For the relaxation of the 2D/3D perovskite heterostructure, we followed the procedure described in Ref. [31].

## V. Models

For all DFTB and DFT calculations, except for the 2D/3D perovskite heterostructure, experimental crystal structures without any geometry optimization were used. The following structures (space groups in parentheses) were considered:

3D perovskites: α-$CsPbI_3$ (Pm-3m) [56], β-$CsPbI_3$ (P4/mbm) [56], γ-$CsPbI_3$ (Pnma) [56], γ-$MAPbI_3$ (Pnma) [68], and α-$FA^*PbI_3$ (Pm-3m) [69] where the symbol * is used to indicate that the FA (formamidinium) organic cation is replaced by a Cs cation at the center of mass to avoid the issue of dynamical disorder of FA at high-temperature phases and the artificial occurrence of a long-range ferroelectric order of the FA molecular dipoles in calculations with periodic boundary conditions.

2D perovskites: References for the structures considered in this study are provided in Table 3 of the main text. In 2D systems, the subscript *n* in the formula denotes the number of inorganic octahedra, related to the thickness of the inorganic layer, within the structure. For example, in $BA_2MA_{n-1}Pb_nI_{3n+1}$, where BA represents butylammonium, *n* specifies the layer count of the inorganic octahedra.

Other systems: We also considered the vacancy ordered perovskite $Cs_4PbI_6$ [70], and the lead- and iodide deficient α-$[HC(NH_2)_2]PbI_3$ perovskites [71,72].

## VI. Energy level and band alignment calculations

To calculate the valence band alignment $\Delta E_\mathrm{v}(3D|2D)$ between 3D and 2D perovskites, we follow the procedure introduced by Wei and Zunger [73], where the band offset is given by:

$$\Delta E_\mathrm{v}(3D|2D) = \Delta E_{\mathrm{v},C'}^{2D} - \Delta E_{\mathrm{v},C'}^{3D} + \Delta E_{C,C'}^{3D/2D} \quad (10)$$

Here,

$$\Delta E_{\mathrm{v},C'}^{3D} = E_\mathrm{v}^{3D} - E_{C'}^{3D} \quad (11)$$

$$\Delta E_{v,C'}^{2D} = E_v^{2D} - E_{C'}^{2D} \quad (12)$$

are the core level (C) to valence band maximum energy separations for pure 3D and 2D perovskites, while

$$\Delta E_{C,C'}^{3D/2D} = E_{C'}^{2D} - E_{C}^{3D} \tag{13}$$

is the difference in core level energy between the 3D and 2D perovskite at the 3D/2D heterostructure. To obtain the unstrained "natural'' offsets, the core to VBM energy difference $\Delta E_{\nu,C'}^{3D}$ and $\Delta E_{\nu,C'}^{2D}$ are calculated for 3D and 2D perovskites at their respective equilibrium (DFT relaxed) lattice constants. The core level difference $\Delta E_{C,C'}^{3D/2D}$ is obtained here from the calculation for the $(3D|2D)$ superlattices. Conduction band offsets can be obtained from the calculated valence band offsets by adding the difference of experimental band gaps. It should be noted that core levels such as Pb 4f and I 3d orbitals are commonly used for core level alignment in experimental and theoretical studies [74-76]. However, these orbitals are not included both in our DFTB and DFT basis, which utilizes the $5s^2 5p^2$ basis for iodine and $6s^2 6p^2$ for Pb. In our investigation, we rely on the I 4s orbital for core-level alignment. Since DFTB employs a frozen core approximation and SIESTA uses norm-conserving pseudopotentials, deep core levels such as the I 4s are not included in their basis sets. Therefore, we obtained the I 4s core energy from a free atom calculation to serve as a stable reference, neglecting the subtle variations in core-level energies caused by environmental effects. This external reference enables us to approximate core-level alignment within our study.

## RESULTS AND DISCUSSION

### DFTB electronic structure for 3D lead iodide perovskites

Following our initial parametrization of DFTB for $CsPbI_3$, we start to test its transferability to 3D LIPs with the general formula $APbI_3$ where A is a monovalent organic or inorganic cation (A = Cs, MA, FA), including tetragonal and cubic perovskite phases. In **Figure 2** and **Table 3**, we compare the results obtained using DFTB with DFT calculations, as well as GW calculations and experimental data, whenever available. First, we calculated the band structure of different perovskite phases [cubic α-phase (Pm-3m), tetragonal β-phase (P4/mbm), and orthorhombic γ-phase (Pnma)] of the inorganic 3D LIP $CsPbI_3$, as shown in **Figure 2 and Table 3**. The results indicate that for all $CsPbI_3$ perovskite phases, our DFTB parameters provide accurate predictions for the band gaps, comparable to GW results and superior to the severely underestimated band gaps typically produced by standard DFT/PBE calculations.

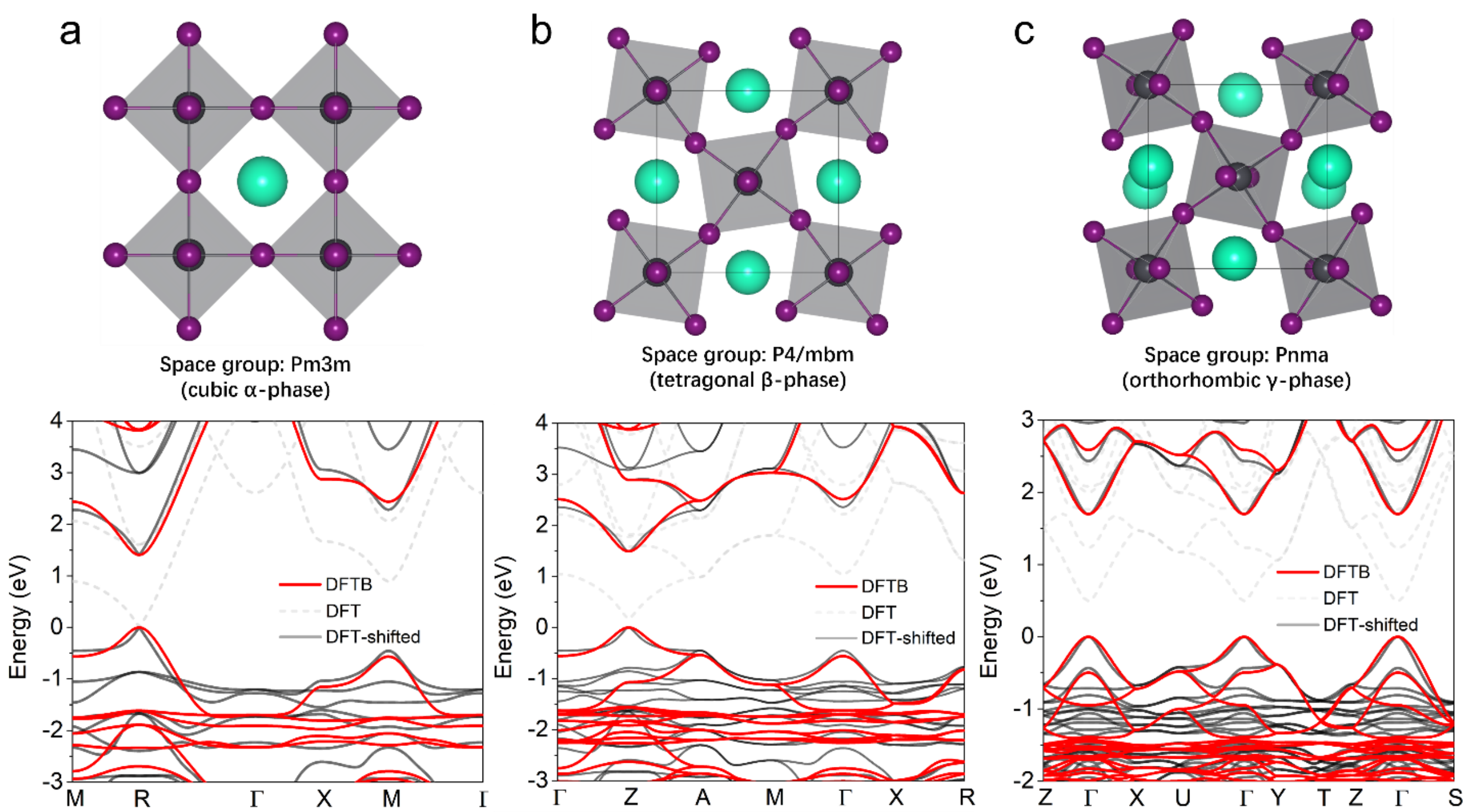

Figure 2. Crystallographic structures of $CsPbI_3$ in (a) cubic, (b) tetragonal, and (c) orthorhombic perovskite phases. The corresponding electronic band structures were calculated using DFTB

including SOC and DFT/PBE with SOC. The top of the valence band is taken as a common energy reference. Each "DFT-shifted" band structure is obtained from a rigid shift of the conduction bands that allows to align CBM computed at both levels of theory. The purple, dark gray, and turquoise spheres denote I, Pb, and Cs atoms, respectively.

We then address the performance of our DFTB parameters for 3D hybrid LIPs. For γ-$MAPbI_3$, the band gap is found to be 1.81 eV in reasonable agreement with experimental values, see **Figure S6** and **Table 2** [57,60]. To further evaluate the accuracy of our DFTB approach with respect to the γ-$MAPbI_3$ electronic structure, **Figure 3** presents a comparison of the band structure obtained from our DFTB parametrization with that from ss-GW as detailed in the Ref. [60]. This comparison shows very good agreement between the DFTB and ss-GW results. Specifically, the discrepancy in the energy difference between the VBM and the lower lying states is 0.16 eV, while the spin-orbit splitting at the CBM shows a minor deviation of 0.04 eV. The effective masses derived from both band structures at the band edges are presented in **Table 2**, with additional data on the effective masses in various directions provided in **Figures S3-S4** and **Tables S4-S5**. The effective masses obtained via DFTB align well with those from ss-GW calculations. The average effective masses over multiple paths in the first Brillouin zone for the γ-$MAPbI_3$ are 0.23 $m_0$ for electrons and 0.28 $m_0$ for holes. The reduced effective mass $\mu$ is 0.13 $m_0$, which agrees well with the experimental value of 0.104 $m_0$ [77,78].

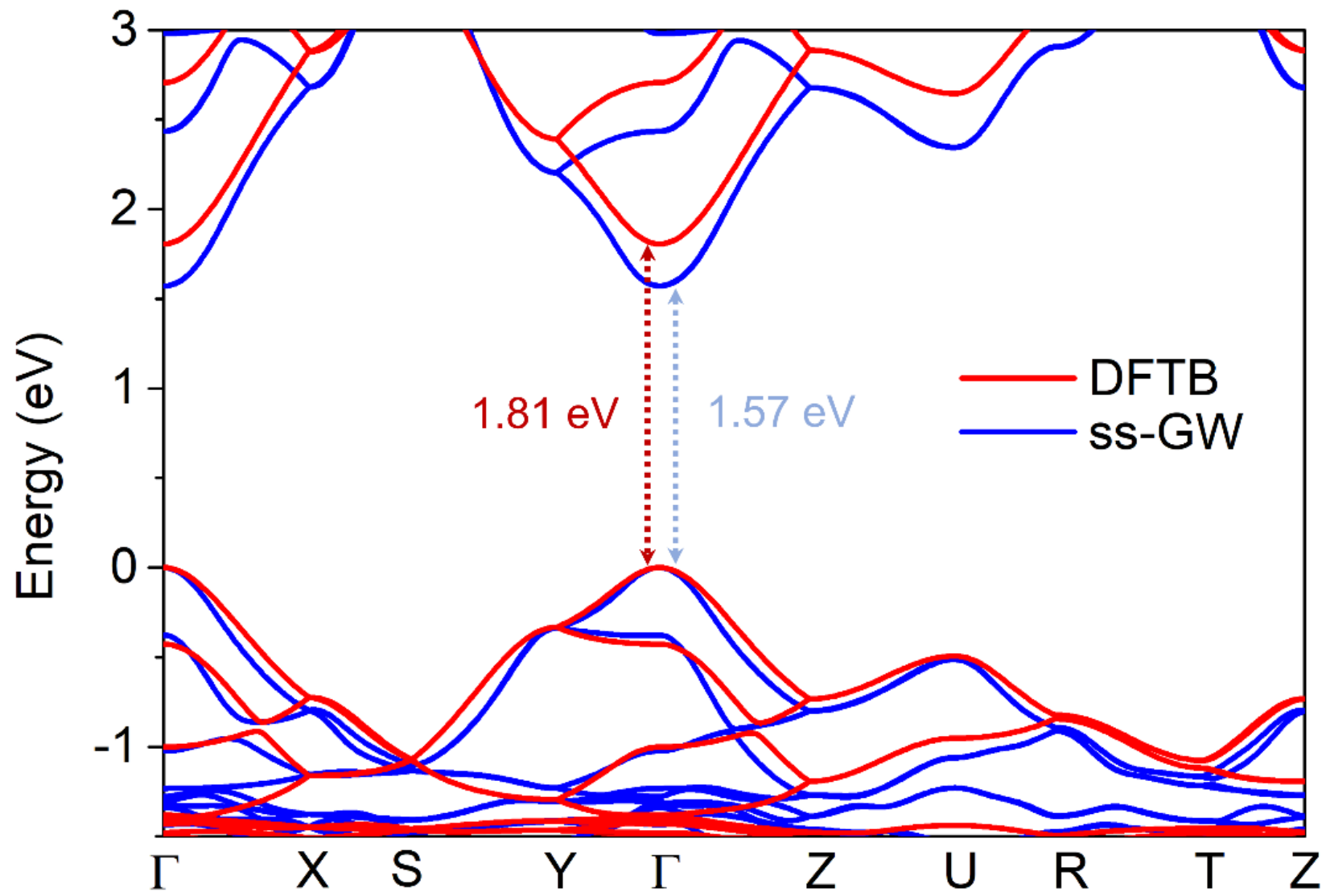

Figure 3. Electronic band structures of γ-$MAPbI_3$ calculated using DFTB, compared with that obtained from self-consistent scissor GW (ss-GW) results of Ref. [60]. The SOC is included in both calculations.

The band gap of the α-$FA^*PbI_3$ is found to be 1.37 eV, which agrees reasonably well with experimental values of 1.48 and 1.52 eV [62,78]. It is important to note that polymorphism (i.e. local disorder) in $FAPbI_3$ is expected not to be as prominent as in other 3D MHPs [79], and as such, is not considered here using a more elaborate theoretical framework [58,59]. The average effective masses over multiple paths in the first Brillouin zone for the α-$FA^*PbI_3$ are 0.18 $m_0$ for electrons and 0.22 $m_0$ for holes. The reduced effective mass $\mu$ amounts to 0.10 $m_0$, which agrees well compared to the experimental result of 0.095 $m_0$ [78]. We highlight that when DFTB is applied to 3D LIPs, there is a noticeable increase in the computed effective masses compared to those calculated using the PBE functional which are severely underestimated [31]. This demonstrates the importance of accurately predicting band gaps to improve in turn the calculations of effective masses.

Table 2. Electronic band gap and effective masses of γ-$MAPbI_3$ from DFTB compared to ss-GW (from Ref. [60]) and experiment (from Refs. [57,60]). $E_g$, $m_e$, $m_h$, and $\mu$ are the band gap, electron, hole, and reduced effective masses, respectively, where $\mu = m_e m_h/(m_e + m_h)$, given in units of the electron rest mass.

| | DFTB | ss-GW | Exp. |
|---|---|---|---|
| $E_g$ (eV) | 1.81 | 1.57 | 1.66/1.65 |
| $m_e$ ($m_0$) | 0.23 | 0.23 | - |
| $m_h$ ($m_0$) | 0.28 | 0.21 | - |
| $\mu$ ($m_0$) | 0.13 | 0.11 | 0.104 |

We also plot the partial density of states (PDOS) of γ-$CsPbI_3$, and γ-$MAPbI_3$ using our DFTB parameters, as shown in **Figure 4**. The results show that the CBM is mainly composed of the Pb 6p orbitals, while the VBM is governed by the I 5p orbitals, which confirms that the character of the states near the VBM and CBM, that dominate the electronic properties of LIPs, is correctly described [41,48]. Furthermore, we also plot the wavefunctions at the VBM and CBM and compare our DFTB results to the DFT results in **Figure 5 and S7**. They show that DFTB successfully captures the atomic orbital character of the band-edge states. It should be noted that SOC is not included for DFT and DFTB wavefunction plots, because the DFTB+ infrastructure does not yet support this feature. We further evaluate the performance of our DFTB parameters for surface studies, by computing the PDOS and wavefunctions for slab models, with details presented in **Figure S8-S9**. When comparing DFTB with DFT for a highly symmetrized 4×4×4 α-$CsPbI_3$ slab, constructed directly from the experimental lattice constant, we observed surface states arising from an artificial Cs CBM state positioned energetically below the actual Pb 6p states. This discrepancy can be addressed by optimizing the atomic positions at the slab surface with DFT, which results in the disappearance of localized wavefunctions on the surface Cs atoms, and recovered the agreement between DFTB and DFT calculations, as can be seen in **Figure S10-S11**.

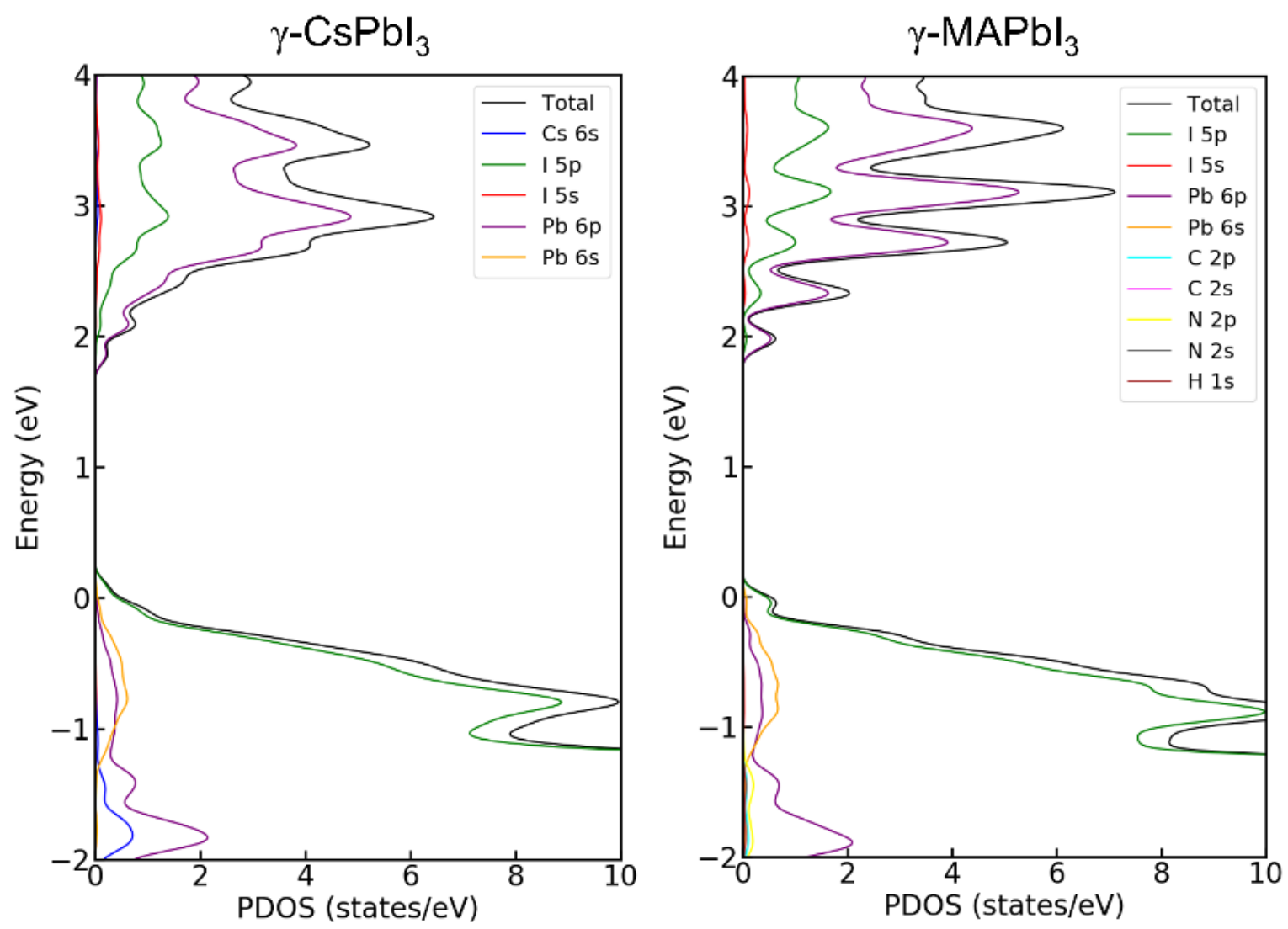

Figure 4. PDOS of 3D LIPs $\gamma$-$CsPbI_3$ and $\gamma$-$MAPbI_3$

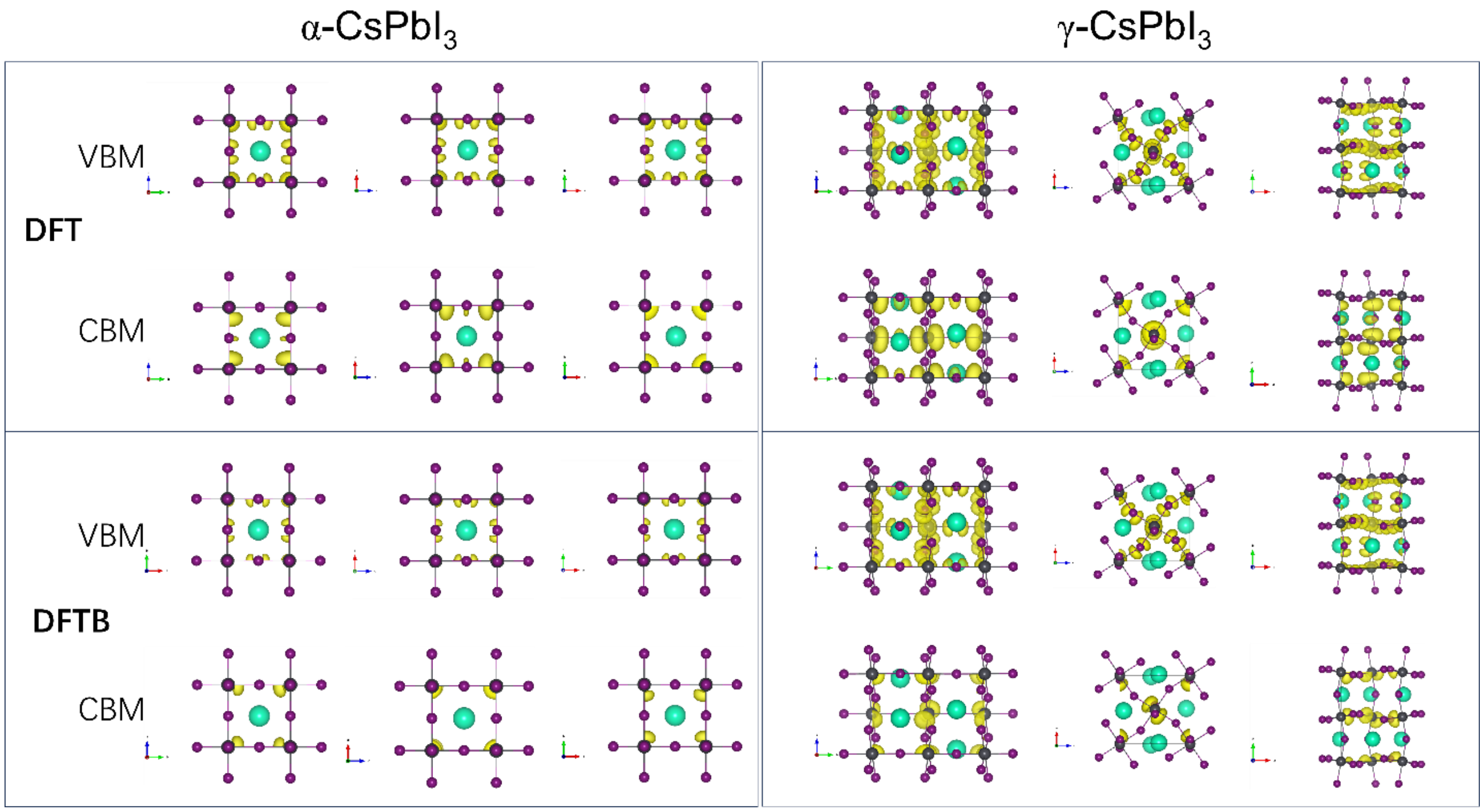

Figure 5. The 3D wavefunction distribution at the VBM and CBM of α-$CsPbI_3$ and γ-$CsPbI_3$ perovskite structures calculated by using DFT and DFTB.

Table 3. Electronic band gaps of 3D LIP obtained at the DFTB, DFT/PBE, and GW@PBE levels, and effective masses obtained with DFTB. The effective masses for electron and hole are the averaged effective masses along the x, y, and z directions. The space group is indicated in parentheses. SOC was included in all calculations. The reduced effective mass $\mu$ is calculated with $\mu = m_e m_h/(m_e + m_h)$. MAE and MAPE stand for mean absolute error and mean absolute percentage error with respect to the experimental values, respectively.

| | Band gap (eV) | | | | Effective mass ($m_0$) | | | |
|---|---|---|---|---|---|---|---|---|
| | This work | This work | Ref. | Ref. | This work | This work | This work | Ref. |
| 3D MHP | DFTB | PBE | GW | Exp. | $m_e$ ($m_0$) | $m_h$ ($m_0$) | $\mu$ ($m_0$) | $\mu_{exp}$ ($m_0$) |
| α-$CsPbI_3$ (Pm-3m) | 1.41 | 0.01 | 1.48 [56] [a] | 1.73 [80] [c]/1.78 [55] [d] | - | - | - | - |
| β-$CsPbI_3$ (P4/mbm) | 1.49 | 0.18 | 1.61 [56] [a] | 1.68 [81] [e] | - | - | - | - |
| γ-$CsPbI_3$ (Pnma) | 1.70 | 0.50 | 2.53 [56] [a] / 1.57 [55] [b] | 1.72 [61,62] [f] | 0.22 | 0.27 | 0.12 | 0.114 [61] [f] |
| γ-$MAPbI_3$ (Pnma) | 1.81 | 0.53 | 1.57 [60] [b] | 1.66 [60]/1.65 [57] | 0.23 | 0.31 | 0.13 | 0.104 [77,78] [g] |
| α-$FA^*PbI_3$ (Pm-3m) | 1.37 | 0.25 | - | 1.48 [62]/1.52 [78] | 0.18 | 0.22 | 0.10 | 0.095 [78] [h] |
| MAE (eV) | 0.09 | 1.19 | 0.12 | | | | 0.012 | |
| MAPE (%) | 5.7 | 73.6 | 7.4 | | | | 11.8 | |

[a] Self-consistent GW calculation

[b] Self-consistent scissors GW calculation

[c] Measured at ⩾608 K

[d] Measured at 623 K

[e] Measured at 483 K

[f] Measured at 2 K

[g] Measured at 155 K–190 K

[h] Measured at 140 K-160 K

The room-temperature structure of prominent hybrid perovskites (first and foremost $MAPbI_3$ and $FAPbI_3$) and the high-temperature phases of $CsPbI_3$ exhibit disordered (polymorphous) networks [79,82,83]. These networks are defined by tilted octahedra and random orientations and/or positions of the A site cations which macroscopically respect the high-symmetry of the structure and minimize the total energy of the system. First-principles calculations using polymorphous networks lead to large renormalizations of the electronic bandgaps and effective masses compared to calculations based on high-symmetry tetragonal and cubic perovskite structures [58,59]. Therefore, evaluating the capability of the DFTB method in accurately capturing this effect of polymorphism is essential. We compared the DFTB-predicted band gaps for both the monomorphous (i.e. high-symmetry perfectly ordered cubic phase) (*hs*) and polymorphous (i.e. locally disordered) (*d*) structures. We generated polymorphous structures of α-$CsPbI_3$ in supercells using the special displacement method [58,59,84] and DFT-PBEsol geometry optimization as described in Refs. [59]. The band structures of polymorphous-$CsPbI_3$ calculated using DFT and DFTB are shown in **Figure S12**. We found that our DFTB parameterization successfully captures the disorder-induced band gap renormalization, as reported in **Table 4**. DFTB performs better than standard DFT and even outperforms popular hybrid functionals such as HSE06, giving results quite comparable to PBE0 and in good agreement with experiment. This highlights the potential of our DFTB parameters to accurately capture effects linked to random local distortions of the perovskite network.

Table 4. Band gap ($E_g$) of high-symmetry (*hs*) monomorphous and disordered (*d*) polymorphous descriptions of the cubic phase of α-$CsPbI_3$. The HSE and PBE0 values are from the Ref. [59]. Calculations of $E_g$ were performed in 2×2×2 supercells.

| | $E_g$ DFTB<br><br>eV | $E_g$ DFT<br><br>eV | $E_g$ HSE<br><br>eV | $E_g$ PBE0<br><br>eV | $E_g$ Exp.<br><br>eV |
|---|---|---|---|---|---|
| *hs*-$CsPbI_3$ | 1.41 | 0.01 | 0.58 | 1.15 | - |
| *d*-$CsPbI_3$ | 1.70 | 0.65 | 1.28 | 1.87 | 1.78 [55] |

**DFTB electronic structure for 2D metal iodide perovskites**

In order to evaluate the transferability of our DFTB parameters to layered (2D) perovskites, we have applied them to a wide range of 2D LIPs, and the results are summarized in **Figure 6** and **Table 5**. We have compared the results obtained using DFTB with DFT/PBE calculations, DFT/HSE06 calculations, and experimental data, whenever literature data are available. We notice that DFTB performs reasonably in predicting the band gaps of 2D compounds compared to reported experimental values. We show for the fictitious model of an undistorted Dion-Jacobson (DJ) phase of $Cs_2PbI_4$, that the electronic band gap is found to be 2.34 eV, which is comparable to the GW value of 2.63 eV [85]. For the room temperature Ruddlesden-Popper (RP) phase $BA_2PbI_4$, the band gap is found to be 2.55 eV and in reasonable agreement with experimental values, although still around 0.3 eV lower as compared to the continuum threshold from magneto-absorption experiments [86]. The reduced effective mass $\mu$ is 0.15 $m_0$, representing an improved prediction compared to experimental values of 0.18 $m_0$ [87] and 0.221 $m_0$ [88], and offering better accuracy than DFT, which yields a lower value of 0.10 $m_0$ [31]. We also report the band gaps and in-plane effective masses of various other 2D LIP. For example, for the multilayered $BA_2MA_{n-1}Pb_nI_{3n+1}$ compounds, the DFTB results show similar agreement with experimental values despite the increase of the number of inorganic octahedral layers in the quantum well [88]. These results suggest that our DFTB parameterization can reasonably capture the quantum confinement effect in 2D perovskites.

It should be noted that due to the fact that the DFTB parametrization primarily tunes the inorganic states (I and Pb states) to adjust the band gap, the band-edge states can be shifted while molecular states are unaffected. This aspect is mostly observed for 2D LIPs with the present DFTB parameterization. In the case of $PEA_2MA_{n-1}Pb_nI_{3n+1}$ and $(GA)MA_nPb_nI_{3n+1}$, the molecular states are indeed spuriously located within the band gap, as illustrated in **Figure S13-S14**. This occurrence can be explained by the layered structure of these materials, which can be viewed as composite materials or heterostructures. In such systems, the inorganic and organic components have distinct band alignments, and the band edges of these components may align differently based on the nature of the molecular spacers [89,90]. One should note, that compression radii for the C, N and H atoms were taken from the *mio-1-1* set [32,91], which was created for describing molecules in vacuum. Adapting those for a better description within the perovskites might remove those artefacts and will be explored in a future work. To address the issue of in-gap molecular states appearing in 2D MHPs, we implemented a strategy involving the substitution of the +1 (+2) positively charged intercalated molecules with one (two) cesium ions ($Cs^+$). This approach has been found to be effective in previous studies for accurately capturing the band structure of organic-inorganic perovskite materials and related compounds [90,92]. By applying this approach, we restored band gaps primarily defined by the inorganic components of the structures, as shown in **Figure S15**. The corrected band gap values are underlined in **Table 5**.

We also plot the PDOS of various 2D LIPs by using our DFTB parameters, see **Figure 7**. The results show that the states at the CBM are mostly composed of the Pb 6p state while those at the VBM are by the I 5p states. This confirms further that the atomic orbital characters of the electronic states near the VBM and CBM, which dominate the electronic properties of 2D LIPs, are successfully captured [93,94]. Furthermore, the localization of the DFTB 3D wavefunctions at the VBM and CBM compare well with DFT results, as shown in **Figure 8**. Similar to the case of 3D LIPs, the SOC is also not included for all DFT and DFTB wavefunction plots.

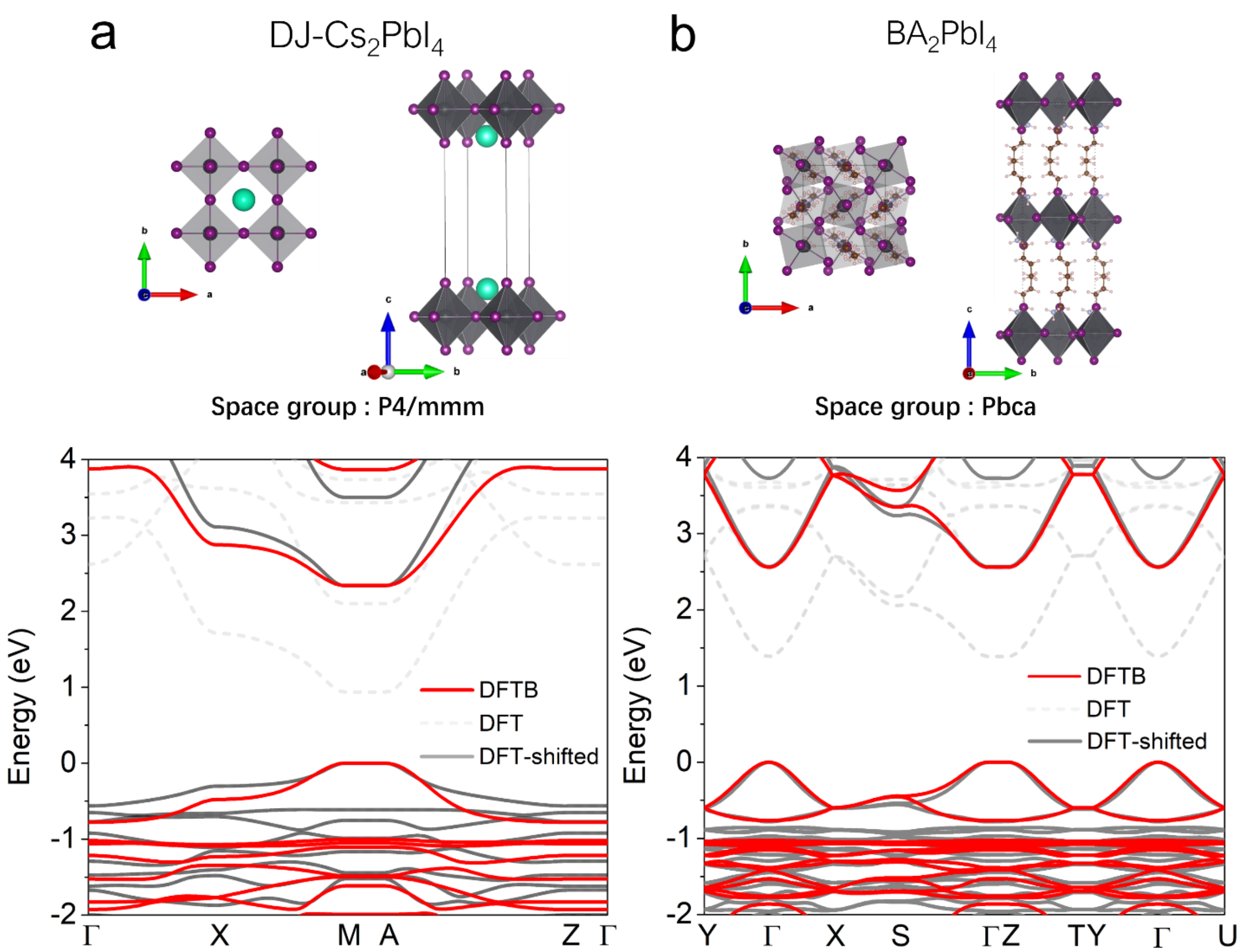

Figure 6. Crystallographic structures of (a) the fictitious 2D model compound DJ-$Cs_2PbI_4$ and (b) and $BA_2PbI_4$. The corresponding electronic band structures were calculated using DFTB including SOC and DFT/PBE with SOC. The top of the valence band is taken as a common energy reference. Each "DFT-shifted" band structure is obtained from a rigid shift of the conduction bands that allows to align CBM computed at both levels of theory. The purple, light blue, brown, light pink, dark gray, and turquoise spheres denote I, N, C, H, Pb, and Cs atoms, respectively.

Table 5. Electronic band gaps of 2D LIPs obtained with DFTB, PBE, HSE06, and in-plane effective masses obtained with DFTB. The space group is indicated in parentheses. SOC was included in all calculations. The reduced effective mass $\mu$ is calculated with $\mu = m_e m_h/(m_e + m_h)$. Underlined data indicates that it was necessary to replace the cation by Cs. The superscripts *opt* and *el* stands for optical and electronic band gaps, respectively. MAE and MAPE stand for mean absolute error and mean absolute percentage error, respectively, with respect to the experimental values.

| | Band gap (eV) | | | | | Effective mass ($m_0$) | | | |
|---|---|---|---|---|---|---|---|---|---|
| | | This work | This work | Ref. | | This work | This work | This work | Ref. |
| 2D MHP | Layer thickness | DFTB | PBE | HSE06 | Exp. | $m_e$ ($m_0$) | $m_h$ ($m_0$) | $\mu$ ($m_0$) | $\mu_{exp}$ ($m_0$) |
| DJ-$Cs_2PbI_4$ | n = 1 [85] | 2.34 | 0.94 | - | 2.63 [85] [a] | - | - | - | - |
| $BA_2MA_{n-1}Pb_nI_{3n+1}$ | n = 1 [95] | 2.55 | 1.28 | 1.96 [96] /2.82 [31] [b] | 2.64 [97]/ 2.80 [86] *el*/ 3.0 [88] *el* | 0.24 | 0.36 | 0.15 | 0.18 [87] / 0.221 [88] [c] |
| | n = 2 [98] | 2.15 | 0.87 | 2.29 [31] [b] | 2.44 [86] *el* | 0.23 | 0.38 | 0.14 | 0.22 [88] [c] |
| | n = 3 [98] | 1.83 | 0.65 | | 2.26 [86] *el* | 0.21 | 0.26 | 0.12 | 0.20 [88] c |
| | n = 4 [98] | 1.82 | 0.64 | | 2.15 [86] *el* | 0.21 | 0.26 | 0.12 | 0.20 [88] c |
| $PEA_2MA_{n-1}Pb_nI_{3n+1}$ | n = 1 [99] | <u>2.41</u> | 1.04 | 2.17 [100] | 2.53 [101] *opt* / 2.87 | 0.22 | 0.30 | 0.13 | 0.091 [102] /0.087 [103] |
| | | | | | [102] *el* | | | | |
| | n = 3 [104] | 1.93 | 0.89 | | 2.10 [105] *opt* / 2.50 [103] *el* | 0.26 | 0.31 | 0.14 | 0.090 [103] |
| $(3AMP)_2MA_{n-1}Pb_nI_{3n+1}$ | n = 1 [106] | 2.37 | 1.02 | | 2.23 [106] *opt* | 0.21 | 0.32 | 0.13 | |
| | n = 2 [106] | 1.41 | 0.39 | | 2.02 [106] *opt* | 0.22 | 0.26 | 0.12 | |
| | n = 3 [106] | 1.48 | 0.30 | | 1.92 [106] *opt* | 0.20 | 0.27 | 0.11 | |
| $(4AMP)_2MA_{n-1}Pb_nI_{3n+1}$ | n = 1 [106] | 2.32 | 1.07 | | 2.38 [106] *opt* | 0.27 | 0.54 | 0.18 | |
| | n = 2 [106] | 1.61 | 0.61 | | 2.17 [106] *opt* | 0.25 | 0.31 | 0.14 | |
| | n = 3 [106] | 1.79 | 0.63 | | 1.99 [106] *opt* | 0.22 | 0.28 | 0.12 | |
| $(GA)MA_nPb_nI_{3n+1}$ | n = 1 [107] | 2.45 | 1.12 | | 2.27 [107] *opt* | 0.22 | 0.42 | 0.14 | |
| | n = 2 [107] | 1.86 | 0.72 | | 1.99 [107] *opt* | 0.23 | 0.27 | 0.12 | |
| | n = 3 [107] | 1.07 | 0.16 | | 1.73 [107] *opt* | 0.20 | 0.25 | 0.11 | |
| $BA_2EA_2Pb_3I_{10}$ | Ref. [108] | 2.04 | 0.99 | | 2.12 [108] *opt* | 0.24 | 0.38 | 0.15 | |
| MAE (eV) | | 0.24 | 1.42 | 1.19 | | | | 0.056 | |
| MAPE (%) | | 11.72 | 65.03 | 15.38 | | | | 37.51 | |

[a] $G_0W_0$ calculation

[b] Exchange $\alpha$ set as 60%

[c] Measured at 4 K

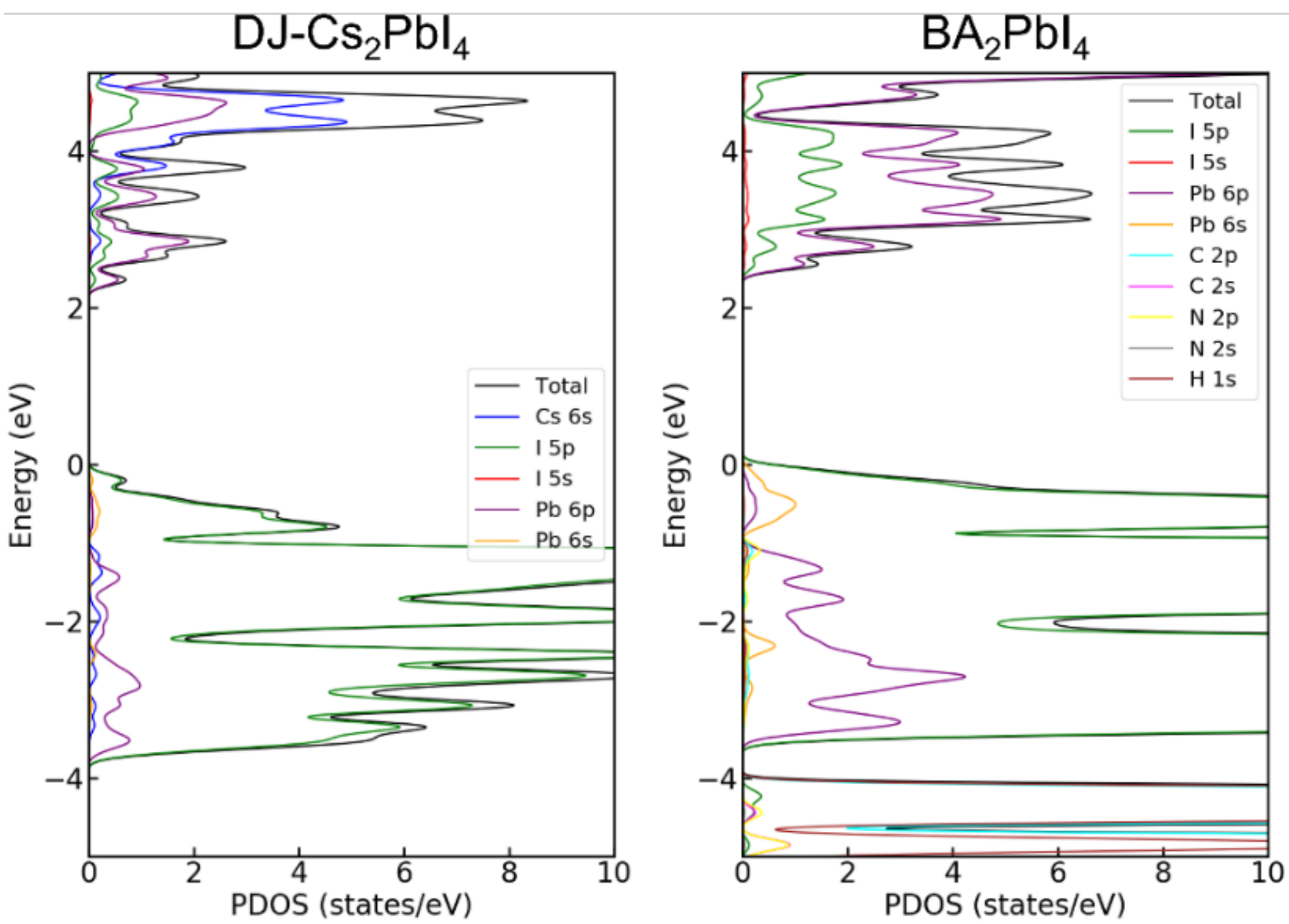

Figure 7. PDOS of 2D LIPs DJ-$Cs_2PbI_4$ and $BA_2PbI_4$.

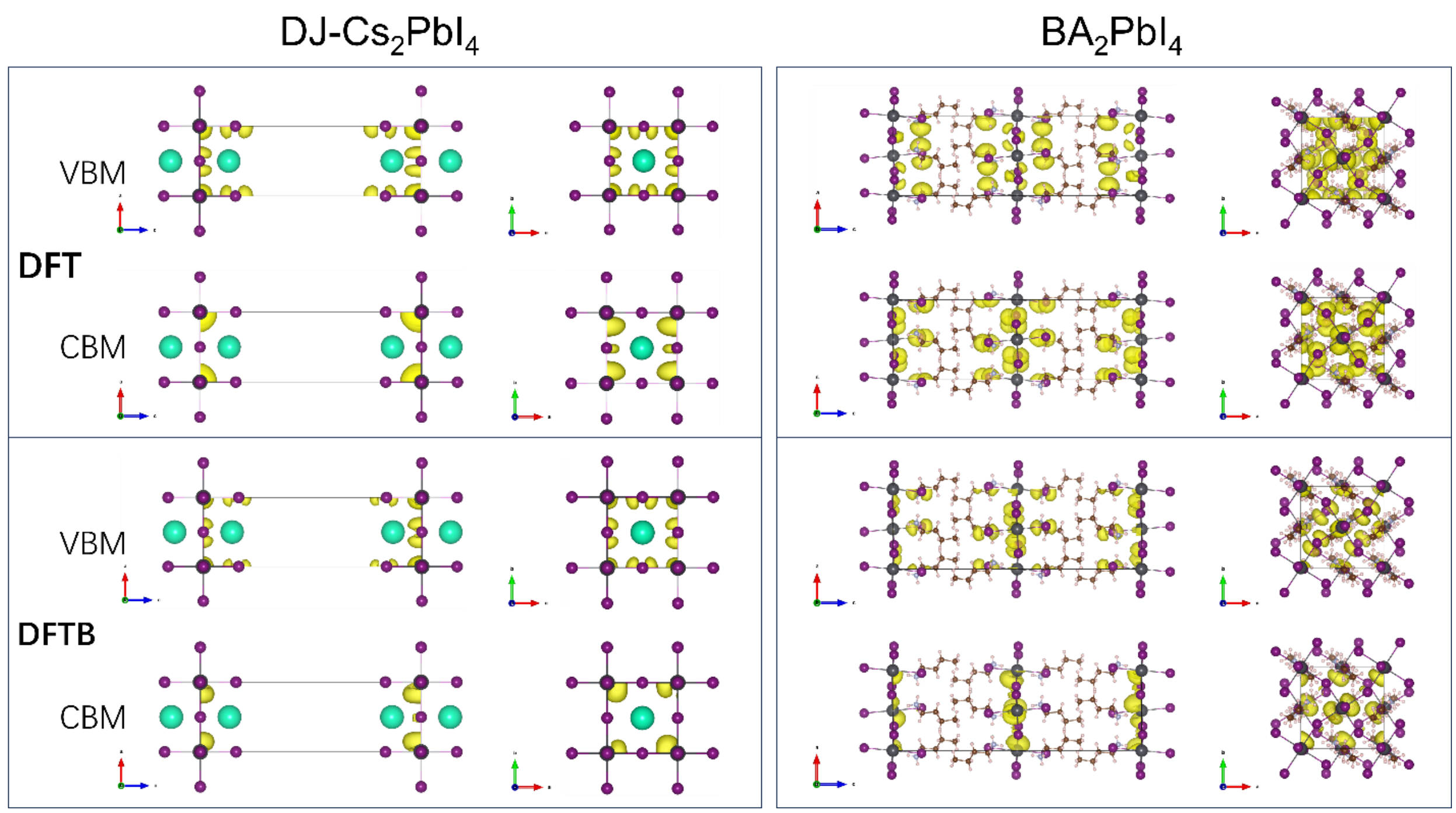

Figure 8. The 3D wavefunction distribution at the VBM and CBM of DJ-$Cs_2PbI_4$ and $BA_2PbI_4$ computed with DFT and DFTB.

**DFTB electronic structure for other perovskites**

In addition to the electronic structures of 3D and 2D LIPs, we investigated other types of perovskite materials, such as the vacancy-ordered perovskite $Cs_4PbI_6$ [70], which has a large experimental optical band gap of 3.38 eV and can be considered as a 0D LIP with isolated $PbI_6^{4-}$ octahedra [109]. Our DFTB parameters return a band gap for $Cs_4PbI_6$ of 3.65 eV, while the DFT predicted value is 2.71 eV (see **Figure S16**). Considering the large exciton binding energy of $Cs_4PbI_6$ as 172 meV [110], our DFTB prediction agrees quite well with experimental measurements. The next perovskite materials we analyze are the lead- and iodide-deficient $FA_{1+x}[Pb_{1-x}I_{3-x}]$ perovskites [71,72]. These architectures can be considered as a bridge between the 3D and 2D perovskites [71], and are attractive for optoelectronic applications because of their reduced lead content, improved stability and the large flexibility of their chemical composition [71,111]. The $FA_{1+x}[Pb_{1-x}I_{3-x}]$, compound with x = 0.2 has an experimental optical band gap of 1.84 eV. As shown in **Figure S16**, the DFTB predicted band gap is 1.44 eV for the crystallographic structure proposed in Ref. [72] for this composition, which is significantly more accurate than the DFT band gap of ~0.92 eV [72]. The plot of the PDOS of $Cs_4PbI_6$ and $FA^*_{1+x}[Pb_{1-x}I_{3-x}]$, x = 0.2 by using our DFTB parameters are shown in **Figure S17**. Overall, our DFTB parameters show strong versatility in predicting the electronic structure across a range of different perovskite materials.

**DFTB for 3D/2D perovskites heterostructures**

We further apply the DFTB parameters to compute band alignments in a 3D/2D LIP heterostructure. Here, we consider the 3D $FA^*PbI_3$ interfaced with the $BA_2PbI_4$ 2D perovskite, as shown in **Figure 9a**. Details for computing the band alignments are given in the Methods section. **Figures 9b-9d** compare the computed band offsets using DFT-PBE (without and with SOC) to DFTB (with SOC) results. The DFT+SOC level of theory is widely known to lead to the closure of the band gap in perovskites [31,112], failing to accurately predict their electronic structure, leading in turn to incorrect band offsets. In contrast, DFT calculations without SOC give a better agreement with experimental band gaps as a result of error compensations [48]. Compared to DFT+SOC, this incidental agreement manifests itself in incorrect band offsets for both the VBM and the CBM. Still, all levels of theory predict a Type-I band alignment. Focusing on the DFTB band offsets, from the alignment of the VBM, the 2D layer presents a barrier to holes collected from the 3D region. Based on our previous discussions on predicting ionization energies [31] and experimental observation [13,113], a smaller barrier (< 0.2 eV) for hole extraction from 2D layer for the 3D/2D perovskite heterostructure is consistently captured by our DFTB parametrization.

For the CBM, a considerable upward shift is observed when comparing the 2D region to the 3D region, a result of quantum confinement that leads to electron blocking. This observation reasonably aligns with experimental findings [13,113]. Therefore, we believe that our computed valence band offsets from DFTB from the heterostructure are reliable. It should be noted that hybrid functional or GW calculations are computationally challenging for the complete semiconductor heterostructure. Alternatively, it was shown for some conventional semiconductor heterostructures [114] that band alignments computed at the DFT-level for a complete heterostructure and combined with GW self-energy corrections for bulk compounds may lead to correct predictions. However, in the present case, such solution would still require to perform GW calculations including SOC for bulk 2D LIPs, with a high computational cost. This example underscores the potential of the DFTB method to accurately predict band alignments in heterostructures, an essential feature for optoelectronic devices, while significantly reducing computational cost compared to standard DFT/GW calculations.

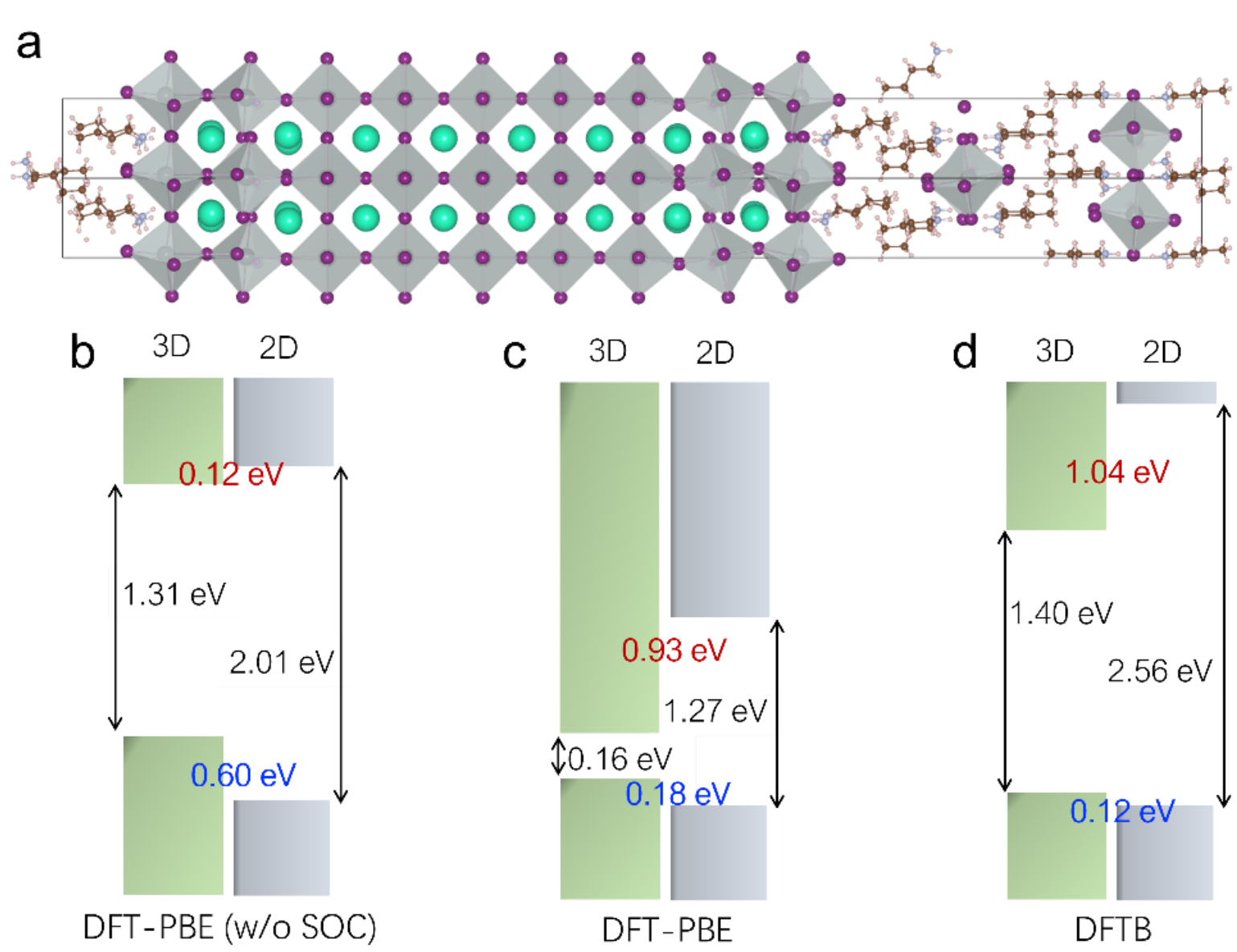

Figure 9. DFTB band alignment computation of a sizeable 2D/3D heterostructure. (a) Constructed 2D/3D $FA^{*}PbI_3/BA_2PbI_4$ heterostructure model. In the heterostructure, both the 2D and 3D regions were terminated with BAI. (b) Computed band alignment using (b) PBE without SOC, (c) PBE with SOC, and (d) DFTB with SOC.

## CONCLUSIONS

In summary, we have developed an electronic parametrization for the DFTB method tailored to lead iodide perovskites (LIPs), with a particular emphasis on aligning with experimental data. This comprehensive assessment of DFTB parameters for calculating essential properties of 3D and 2D LIPs includes electronic band gaps, effective masses, and heterostructure band alignments, significantly enhancing predictive accuracy and efficiency. Our DFTB approach aims to achieve transferability across various perovskite materials and brings orders of magnitude improvement in the computational efficiency compared to DFT. This efficiency, combined with the ability to produce band gaps and effective masses close to experimental values, underscores the advantages of the DFTB approach. Their versatility allows the study of other low-dimensional perovskite nanostructures, including vacancy-ordered and lead- and iodide-deficient perovskites, as well as energy level alignments in 3D/2D perovskite heterostructures. Future efforts could focus on repulsive parametrization to complete the DFTB parameters, enabling molecular dynamics simulations and facilitating the modeling of locally disordered structures in larger supercells. Moreover, expanding these DFTB parameters to a broader array of perovskite compositions, including lead-free and mixed-halide perovskites ubiquitous in solar cell applications, would be highly desirable.

## ASSOCIATED CONTENT

**Supplemental Material**. A detailed list of the contents of each file provided as Supplemental Material is included. The data and Slater-Koster files supporting the findings of this study are available within the paper and can be downloaded from http://www.dftb.org/parameters for use by the scientific community.

**Corresponding Authors**

***E-mail:** Simon.thebaud@insa-rennes.fr, jacky.even@insa-rennes.fr, claudine.katan@univ-rennes.fr

## ACKNOWLEDGMENT

We thank Marina R. Filip and Feliciano Giustino for kindly sharing raw data of the ss-GW band structure for our DFTB parametrization. The work at institute FOTON and ISCR acknowledges funding from the M-ERA.NET project PHANTASTIC (R.8003.22), and the European Union's Horizon 2020 program, through an Innovation Action under Grant Agreement No. 861985 (PeroCUBE) and a FET Open research and innovation action under the Grant Agreement No. 899141 (PoLLoC). M.Z. acknowledges funding by the European Union (project ULTRA-2DPK / HORIZON-MSCA-2022-PF-01 /Grant Agreement No. 101106654). Views and opinions expressed are however those of the authors only and do not necessarily reflect those of the European Union or the European Commission. Neither the European Union nor the granting authority can be held responsible for them. J.E. acknowledges financial support from the Institut Universitaire de France. S.T acknowledges financial support from Rennes Metropole under the AIS grant. G.V. acknowledges funding from the Agence Nationale pour la Recherche through the CPJ program and the SURFIN project (ANR-23-CE09-0001). The work was granted access to the

HPC resources of TGCC/CINES under the allocation 2023-A0140911434 and 2024-A0160911434 made by GENCI. We further acknowledge access to computing facilities provided by the Institut National des Sciences Appliquées de Rennes (INSA Rennes) and the Institut des Sciences Chimiques de Rennes (ISCR). T. v. d. H. acknowledges financial support from the German Research Foundation (DFG) through Grant No. FR2833/76-1. C.R.L.-M. acknowledge financial support from the German Research Foundation (DFG) through Grant No. FR 2833/82-1. C.R.L.-M. extends special thanks to A.F. for the invitation to the Université de Rennes in 2022, where he first met C.K., J.E., and S.T., marking the beginning of this collaboration between the groups in Rennes and Bremen.

**TOC**

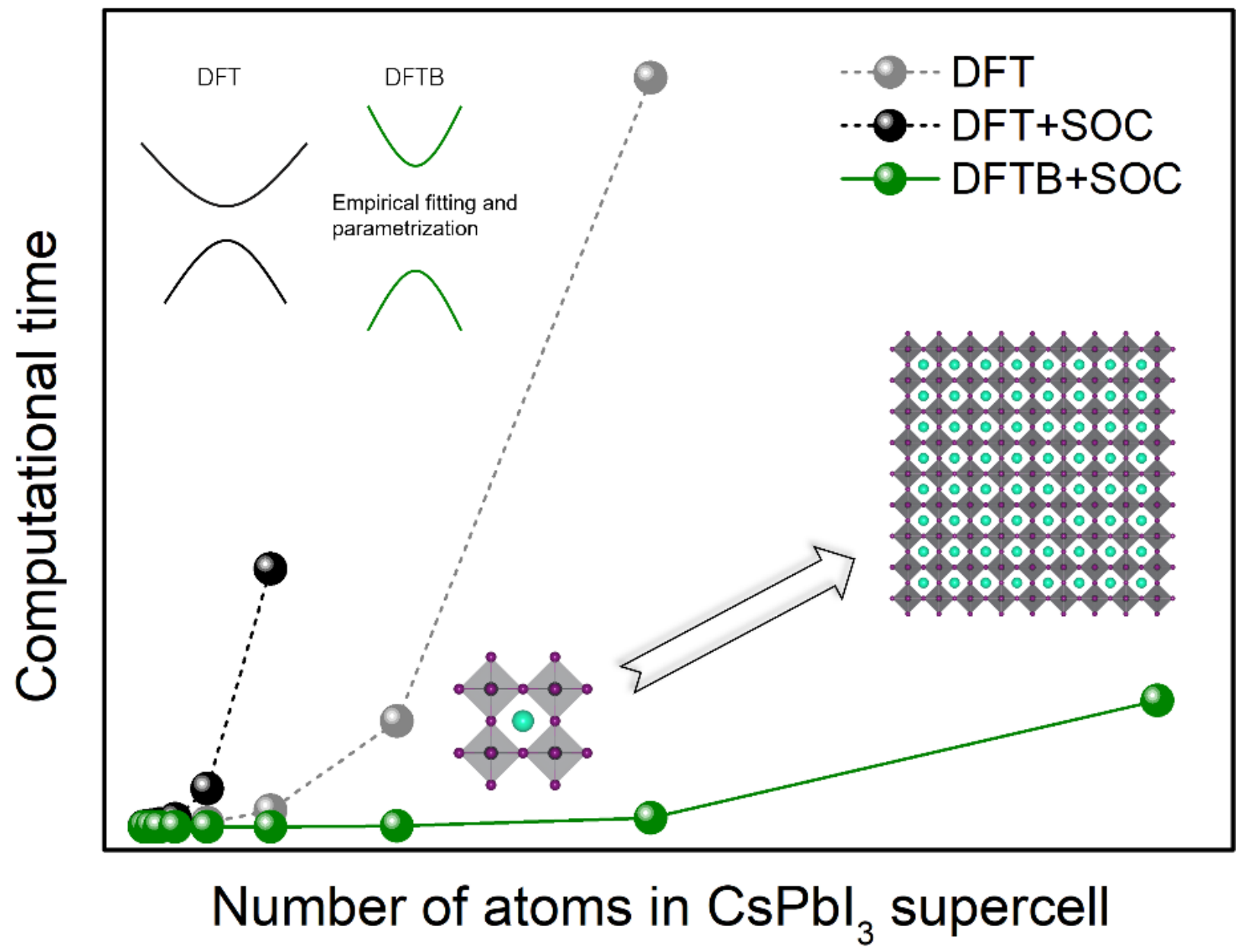

DFT
DFTB
Empirical fitting and parametrization
DFT
DFT+SOC
DFTB+SOC
Computational time
Number of atoms in CsPbI3 supercell